# The Electromagnetic Fields of a Spinning Spherical Shell of Charge

Stanislaw Olbert and John W. Belcher
Department of Physics
Massachusetts Institute of Technology

**Abstract**

We consider the problem of a rotating charged spherical shell of radius *a* carrying an axially symmetric distribution of charge. We give the interior and exterior solutions to this problem for arbitrary zonal dependence of the surface charge distribution and for arbitrary time dependence of the rotation rate $\Omega(t)$, assuming $\Omega a \ll c$. Surprisingly, these general solutions involve the rotation rate and its time integrals, and do not depend on the time derivatives of the rotation rate. We present numerical solutions for various distributions of surface charge, and we consider the case where the sphere instantaneously goes from rest to rotating at a constant rate. For a time $2a/c$ after this instantaneously spin-up, the sphere radiates even though it is rotating at a constant rate. In this time interval, the sphere radiates away an amount of energy exactly equal to the magnetic energy stored in the static fields a long time after the sphere has been spun up. Animations of our solutions can be found at ://web.mit.edu/viz/spin .

# 1 Introduction

Studies of the electromagnetic fields generated by very rapidly changing electric currents are difficult because numerical solutions involve extensive calculation. It is easy to state that **E** and **B** fields are derivable from the vector potential **A**, write down the integral expression for **A** due to a given source current, and state that the integrand must be evaluated at the retarded time. However, even for current distributions with relatively simple geometries (with the exception of infinite planes), these integrals turn out to be non-elementary, not expressible in terms of any known functions, and thus only obtainable by lengthy numerical calculations. Consider, for example, a current changing in a circular loop. If the current changes rapidly compared to the speed of light transit time across the loop, the vector potential is not expressible by any known elementary or non-elementary function. Only in the limit of quasi-stationary currents (that is, for currents which change on a time scale $T$ such that the distance $cT$ may be treated as infinite compared to the radius of the loop) can one express the electromagnetic fields *everywhere* in space in terms of known functions, and even then one deals with various kinds of elliptical integrals.

As a result of these difficulties, we tend to avoid in the classroom detailed discussions of the fields associated with rapidly changing transient currents. Pedagogically, however, we would like a student to be able to observe a change in current in one part of a system causing changes in the local fields, and to observe these field changes as they propagate at the speed of light to other parts of the system. In the quasi-stationary approximation this is impossible, because the resultant changes in the field appear everywhere in the system instantaneously, even though clearly these field changes must propagate at finite speeds. Ideally one would like to show students actual examples where they can observe this finite speed of propagation of the field changes.

In the case of infinite geometries, the case of infinite parallel current sheets changing arbitrarily in time can be treated with ease, which is surprising considering the complexity in any other dimension. For the infinite planar problem, the fields can be written in terms of the time dependence of the planar surface currents at properly retarded times [1]. For an infinite current sheet in the $yz$ plane located at $x = 0$, carrying a current per unit length $\kappa(t)\hat{\mathbf{y}}$, the fields are given simply by

$$\mathbf{E}(x,t) = -\hat{\mathbf{y}}\,\frac{c\mu_o}{2}\kappa\left(t - \frac{|x|}{c}\right)$$

$$\mathbf{B}(x,t) = -\hat{\mathbf{z}}\,\frac{\mu_o}{2}\kappa\left(t - \frac{|x|}{c}\right)sign(x) \tag{1}$$

An infinitely-long solenoid is also treatable [2], albeit with much more difficulty. For this case, to obtain the solution for an arbitrary time dependence of surface currents, the problem must first be solved for the Fourier analyzed time situation, and then the solution for arbitrary time-dependence in the current found using the inverse Fourier transform.

In the general non-infinite three dimensional case, this is also the standard approach, in that we first assume harmonic time dependence, introduce vector spherical harmonics, and then treat arbitrary time dependence using Fourier transforms [3].

We have been searching for non-planar source current configurations that are more tractable numerically than the general non-infinite three dimensional case above, and also illuminating pedagogically, in that we can observe finite transit time effects in non-planar geometries. To observe transit time effects we must of course return from Fourier transform space and consider solutions explicitly in the time domain. The question we address here is the following. Is there a finite, spatially confined current system simple enough to be treated analytically in the time domain which is also useful for the pedagogical purposes outlined above? The answer to this question is yes.

We present here a method for finding the fields of a spinning spherical shell of radius $a$ carrying an axially symmetric surface charge, valid for an arbitrary time dependence of its rotation rate $\Omega(t)$, assuming only that $\Omega a << c$. The final expressions for our fields can be evaluated in the time domain using relatively simple numerical techniques. These expressions are not as simple as are those for an infinite plane, but they can be evaluated numerically very rapidly. We give expressions for the fields both inside and outside the spherical shell. Our solutions do not require us to first Fourier analyze $\Omega(t)$, nor do they require that $\Omega(t)$ have any "nice" properties (for example, $\Omega(t)$ can be discontinuous).

Before embarking on this exposition, we note that a similar solution for a uniformly charged spinning sphere has been given in the region outside the sphere by Daboul and Jensen [4]. However these authors did not give the solution in the region $r < a$, and it is not clear how to extend their method (which uses a contour integration in the complex plane) to surface charge distributions depending on the angle from the spin axis of the sphere. In addition, in an unpublished manuscript, Vlasov [5] also gives the solution for a uniformly charged spinning sphere in a different but equivalent form to that given in [4], valid both inside and outside the sphere. We discuss the form of these two solutions in more detail below (see Eq.(49)). In contrast to these authors, the solutions that we give here are valid for any dependence of the surface charge density on the angle from the spin axis of the sphere, and our solutions are expressed in a very different, although equivalent, form for the constant surface charge case.

## 2   The problem

We state the problem for a general zonal dependence of the charge distribution. We use the polar coordinates ($r, \theta, \phi$) and derive the differential equation for the azimuthal component of the vector potential **A**, which is the only non-vanishing component in the case of axial symmetry. The general differential equation for **A** is

$$-\nabla \times \nabla \times \mathbf{A} - \frac{1}{c^2}\frac{\partial^2 \mathbf{A}}{\partial^2 t} = -\mu_o \mathbf{J} \tag{2}$$

The equation for the $\phi$-component is

$$\frac{1}{r}\frac{\partial^2}{\partial r^2}(rA_\phi) + \frac{1}{r^2}\frac{\partial}{\partial \theta}\left[\frac{1}{\sin\theta}\frac{\partial}{\partial \theta}(\sin\theta A_\phi)\right] - \frac{1}{c^2}\frac{\partial^2}{\partial t^2}A_\phi = -\mu_o j_\phi \tag{3}$$

The current density is given by

$$j_\phi(r,\theta,t) = \frac{\delta(r-a)}{4\pi r} Q\Omega_0 \tilde{S}^{(0)}(t)\Lambda_l(\theta) \tag{4}$$

where $\Omega_0$ is the maximum angular velocity of the shell and $\tilde{S}^{(0)}(t)$ and $\Lambda_l(\theta)$ are some suitably chosen – but otherwise arbitrary – functions of time and polar angle, respectively, with $\Omega(t) = \Omega_0 \tilde{S}^{(0)}(t)$. For spherical geometry, the $\theta$-dependence of $A_\phi$ is separable. That is, we can put

$$A_\phi(r,\theta,t) = A_0 \tilde{A}_l(r,t)\Lambda_l(\theta) \tag{5}$$

where $A_0$ is a constant chosen to render $\tilde{A}_l(r,t)$ and $\Lambda_l(\theta)$ dimensionless

$$A_0 = \frac{\mu_o}{4\pi}Q\Omega_0 \tag{6}$$

and $\Lambda_l(\theta)$ obeys the following ordinary differential equation

$$\frac{d}{d\theta}\left[\frac{1}{\sin\theta}\frac{d}{d\theta}(\sin\theta\, \Lambda_l)\right] = -l(l+1)\Lambda_l(\theta) \tag{7}$$

In general, the value of $l(l+1)$ can be an arbitrary constant. In what follows, we shall limit our discussion to integer $l$'s. For this case the solution to Eq. (7) is

$$\Lambda_l(\theta) = \sin\theta\frac{dP_l(\cos\theta)}{d(\cos\theta)} \tag{8}$$

where $P_l$ is the Legendre polynomial of order $l$. Note that, except for the phase convention, $\Lambda_l(\theta)$ is one of the associated Legendre functions $P_l^m(\cos\theta)$, that is, if $x = \cos\theta$,

$$\Lambda_l(\theta) = -P_l^1(x) = -\sqrt{1-x^2}\,\frac{dP_l(x)}{dx} \tag{9}$$

This implies that the $\Lambda_l$'s form a complete set of orthogonal functions. Specifically, for $m = 1$, we have

$$\int_{-1}^{1} P_{l'}^1(x) P_l^1(x) dx = \frac{2l(l+1)}{2l+1}\delta_{l'l} \tag{10}$$

The factorized representation in Eq.(5) leads then to the following equation for $\tilde{A}$ for a specific value of $l$:

$$\frac{1}{r}\frac{\partial^2}{\partial r^2}(r\tilde{A}_l) - \frac{l(l+1)}{r^2}\tilde{A}_l - \frac{1}{c^2}\frac{\partial^2}{\partial t^2}\tilde{A}_l = -\frac{\delta(r-a)}{r}\tilde{S}^{(0)}(t) \tag{11}$$

We define the four times $t_a$, $t_r$, $t_<$ and $t_<$ as

$$t_a = \frac{a}{c} \qquad t_r = \frac{r}{c} \tag{12}$$

and

$$t_< = \min(t_r, t_a); \qquad t_> = \max(t_r, t_a) \tag{13}$$

To shorten algebraic manipulations, we replace the reduced vector potential function $\tilde{A}_l$ (see Eq. (5)) by a more convenient function, $f_l$ (" the reduced flux function"), defined by

$$f_l(t_>, t_<, t) = t_r \tilde{A}_l(t_r, t_a, t) \tag{14}$$

Note that $f_l$ has dimensions of time, since $\tilde{A}_l$ is dimensionless. With this notation, the quantity $f_l$ obeys the following equation.

$$\frac{\partial^2 f_l}{\partial t_r^2} - \frac{l(l+1)}{t_r^2} f_l - \frac{\partial^2 f_l}{\partial t^2} = -\delta(t_r - t_a)\tilde{S}^{(0)}(t) \tag{15}$$

The exact solution of this equation for any given $\tilde{S}^{(0)}(t)$ is the subject of this paper.

### 3    The solution for arbitrary order $l$

We now solve Eq. (15) by Laplace transforming with respect to time and then inverting. We could use Fourier transforms as well, but we prefer Laplace transforms because of their more compact and pedagogically more instructive appearance. We assume that the Laplace transform of the function $\tilde{S}^{(0)}(t)$ (cf. Eq. (4)) for the time dependence of the spin rate has been worked out and is available. That is, we have the function $\bar{S}^{(0)}(s)$ defined by

$$\bar{S}^{(0)}(s) = \int_0^\infty \tilde{S}^{(0)}(t) e^{-st} dt \tag{16}$$

Introducing the image function (the Laplace transform) of $f_l$

$$\bar{f}_l(t_r, t_a, s) = \int_0^\infty f_l(t_r, t_a, t) e^{-st} dt \tag{17}$$

we have from Eq.(15)

$$\left[ \frac{d^2}{dt_r^2} - \left( \frac{l(l+1)}{t_r^2} + s^2 \right) \right] \bar{f}_l = -\delta(t_r - t_a) \bar{S}^{(0)}(s) \tag{18}$$

Eq. (18) is analogous to the familiar Fourier-transformed equation for spherical waves if we replace the variable $t_r$ by $r$ and $s$ by $i\omega/c = ik$. To be more precise, if we change the independent variable $t_r$ to $z = i\omega t_r$, the homogeneous part of Eq.(18) is referred to in the classical literature as the Riccati-Bessel equation [6]. If $j_l(x)$ and $h_l^{(1)}(x)$ are the spherical Bessel and Hankel functions in conventional notation [7], then solutions to the Riccati-Bessel equation are $z j_l(z)$ and $z h_l^{(1)}(z)$ respectively. We define

$$\bar{j}_l(x) \equiv (-i)^l j_l(ix) \qquad \bar{h}_l(x) \equiv -i^l h_l^{(1)}(ix) \tag{19}$$

Adopting this notation, one can find in the literature the following well known solution to Eq. (18)

$$\bar{f}_l(t_>, t_<, s) = C(s) \, t_r \, \bar{j}_l(t_< s) \bar{h}_l(t_> s) \bar{S}^{(0)}(s) \tag{20}$$

The factor $C(s)$ in Eq. (20) must be determined from the boundary conditions across the surface of the sphere. After we derive the expression for the **B**-field, we can show from the boundary conditions that (see Eq. (50) below)

$$C(s) = s t_a \tag{21}$$

Using Eqs. (20) and (21), we can invert our Laplace transform using the standard inversion formula

$$f_l(t_>,t_<,t) = \frac{1}{2\pi i}\int_{c-i\infty}^{c+i\infty} \bar{f}_l(t_>,t_<,s)e^{ts}ds$$

$$= \frac{1}{4\pi i}\int_{c-i\infty}^{c+i\infty} ds\, e^{ts}\, st_a t_r\, \bar{j}_l(t_<s)\bar{h}_l(t_>s)\bar{S}^{(0)}(s) \quad (22)$$

To perform the integration in Eq.(22), first note that the functions $\bar{j}_l$ and $\bar{h}_l$ in Eq. (20) are closely related to the Modified Spherical Bessel Functions of the first and third kind, $I_{l+\frac{1}{2}}(x)$ and $K_{l+\frac{1}{2}}(x)$, respectively. With the help of items 10.1.9 and 10.2.15 in [8], we can write

$$\bar{h}_l(x) = \frac{e^{-x}}{x}\sum_{m=0}^{l}\frac{\gamma_{l,m}}{(2x)^m} \quad (23)$$

where the coefficient $\gamma_{l,m}$ is given by

$$\gamma_{l,m} = \frac{(l+m)!}{m!(l-m)!} \quad (24)$$

Our symbol $\gamma_{l,m}$ is identical in meaning with the Hankel symbol $(n,m)$ defined as

$$(n,m) = \frac{\Gamma(\tfrac{1}{2}+n+m)}{m!\,\Gamma(\tfrac{1}{2}+n-m)} \quad (25)$$

if we put $n = l + \tfrac{1}{2}$. With this notation one can then readily show that

$$\bar{j}_l(x) = \frac{(-1)^l}{2x}\sum_{m=0}^{m=l}\frac{\gamma_{l,m}}{(2x)^m}\left[(e^x - e^{-x})\frac{(1+(-1)^{l+m})}{2} - (e^x + e^{-x})\frac{(1+(-1)^{l+m+1})}{2}\right] \quad (26)$$

Inserting Eqs.(23) and (26) into Eq.(22), one finds that each individual term on the right-hand side of Eq. (22) has the following mathematical structure

$$\frac{1}{2\pi i}\int_{c-i\infty}^{c+i\infty}\frac{1}{s^m}\bar{S}^{(0)}(s)e^{s(t-t_> \pm t_<)}ds \quad (27)$$

This implies that we are dealing with a series of consecutive integrals of the function $\tilde{S}(t)$. Starting with $m = 1$, we define our first serial integral $\tilde{S}^{(1)}(t)$ by

$$\tilde{S}^{(1)}(t) \equiv \int_0^t \tilde{S}^{(0)}(t')dt' = \frac{1}{2\pi i}\int_{c-i\infty}^{c+i\infty}\frac{1}{s}\bar{S}^{(0)}(s)e^{ts}ds \quad (28)$$

and, in succession, for higher $m$'s,

$$\tilde{S}^{(m)}(t) = \int_0^t \tilde{S}^{(m-1)}(t')\,dt' = \frac{1}{2\pi i} \int_{c-i\infty}^{c+i\infty} \frac{1}{s^m} \overline{S}^{(0)}(s) e^{ts}\,ds \tag{29}$$

With this notation we see that Eq. (27) can be written simply as $\tilde{S}^{(m)}(t - t_> \pm t_<)$ and that the solution for $f_l(t_>, t_<, t)$ is a linear composition of $\tilde{S}^{(m)}(t - t_> \pm t_<)$. After a straight forward algebraic effort one finds that $f_l(t_>, t_<, t)$ may be cast into the following form

$$f_l(t_>, t_<, t) = \frac{1}{2} \sum_{k=0}^{l} \sum_{m=0}^{l} \frac{\gamma_{l,k}\gamma_{l,m}}{(2t_>)^k (2t_<)^m} \left[ (-1)^m \tilde{S}^{(k+m+1)}(t_+) + (-1)^{l+1} \tilde{S}^{(k+m+1)}(t_-) \right] \tag{30}$$

where

$$t_\pm = t - t_> \pm t_< \tag{31}$$

Eq. (30) is our exact solution to Eq.(15) for $f_l$ in the time domain, in terms of the serial integrals of $\tilde{S}^{(0)}(t)$ as defined by Eq. (29).

## 4  Expressions for E and B

Given $f_l$, we now derive expressions for the time-dependent transient electric and magnetic fields. Using spherical coordinates, we have in general for our geometry that

$$E_\phi(r,\theta,t) = -\frac{\partial A_\phi(r,\theta,t)}{\partial t} \tag{32}$$

With the help of $\mathbf{B} = \nabla \times \mathbf{A}$, we see that $B_\phi(r,\theta,t) = 0$ and that

$$B_r(r,\theta,t) = \frac{1}{r\sin\theta} \frac{\partial}{\partial \theta}\left(\sin\theta A_\phi\right) \tag{33}$$

$$B_\theta(r,\theta,t) = -\frac{1}{r}\frac{\partial}{\partial r}\left(rA_\phi\right) \tag{34}$$

The above list of field components do not contain $E_r$ and $E_\theta$ because these components are time independent and thus are not of direct relevance to our discussion. We will quote the time independent components for $l = 1$ when we consider that case in detail below. We introduce the following symbolic abbreviations to simplify our notation. We define

$$E_0 = \frac{A_0}{t_a} = \frac{Q\Omega_o}{4\pi\varepsilon_o ac} = \frac{Q}{4\pi\varepsilon_o a^2}\left(\frac{\Omega_o a}{c}\right) \tag{35}$$

and

$$B_o = \frac{E_0}{c} = \frac{\mu_0 Q\Omega_0}{4\pi a} \tag{36}$$

For future reference, we note that for the case of a uniformly charged spherical shell rotating at a constant rate $\Omega_0$, the magnetic field inside the sphere is parallel to the spin axis with magnitude $\frac{2}{3}B_0$ and that the magnetic dipole moment of the shell is given by

$$m_0 = \frac{1}{3}a^2\Omega_0 Q \tag{37}$$

We now introduce the "reduced" (dimensionless) expressions for $E_\phi$, $B_r$, and $B_\theta$

$$\tilde{E}_{l,\phi}(t_r,t_a,t) = \frac{E_\phi(r,\theta,t)}{E_0 \Lambda_l(\theta)} \tag{38}$$

$$\tilde{B}_{l,r}(t_r,t_a,t) = \frac{B_r(r,\theta,t)}{l(l+1)B_0 P_l(\cos\theta)} \tag{39}$$

$$\tilde{B}_{l,\theta}(t_r,t_a,t) = \frac{B_\theta(r,\theta,t)}{B_0 \Lambda_l(\theta)} \tag{40}$$

where $\Lambda_l$ is given by Eq.(9) and the $P_l$'s are the Legendre polynomials. With these abbreviations we readily find that Eqs. (38), (39) and (40) can be written as follows

$$\tilde{E}_{l,\phi}(t_r,t_a,t) = -\frac{t_a}{t_r}\frac{\partial f_l(t_>,t_<,t)}{\partial t} \tag{41}$$

$$\tilde{B}_{l,r}(t_r,t_a,t) = \frac{t_a}{t_r^2} f_l(t_>,t_<,t) \tag{42}$$

$$\tilde{B}_{l,\theta}(t_r,t_a,t) = -\frac{t_a}{t_r}\frac{\partial f_l(t_>,t_<,t)}{\partial t_r} = -\frac{t_a}{t_r}\begin{cases} \dfrac{\partial f_l(t_>,t_<,t)}{\partial t_<} & \text{for } t_r < t_a \\[2mm] \dfrac{\partial f_l(t_>,t_<,t)}{\partial t_>} & \text{for } t_r \geq t_a \end{cases} \tag{43}$$

The explicit representation of $f_l(t_>,t_<,t)$ given in Eq. (30) makes it easy to work out the derivatives of $f_l$ with respect to any of the three arguments $t_>$, $t_<$ and $t$ for any order of $l$, and thus to find explicit expressions for our field components. In particular, we have that

$$\frac{\partial f_l(t_>,t_<,t)}{\partial t} = \frac{1}{2}\sum_{k=0}^{l}\sum_{m=0}^{l}\frac{\gamma_{l,k}\gamma_{l,m}}{(2t_>)^k(2t_<)^m}\left[(-1)^m\tilde{S}^{(k+m)}(t_+) + (-1)^{l+1}\tilde{S}^{(k+m)}(t_-)\right] \quad (44)$$

$$\frac{\partial f_l}{\partial t_<} = \frac{1}{2}\sum_{k=0}^{l}\sum_{m=0}^{l+1}\frac{\gamma_{l,k}\gamma_{l,m}}{(2t_>)^k(2t_<)^m}\left[(-1)^m\tilde{S}^{(k+m)}(t_+) - (-1)^{l+1}\tilde{S}^{(k+m)}(t_-)\right] \quad (45)$$

and

$$\frac{\partial f_l}{\partial t_>} = -\frac{1}{2}\sum_{k=0}^{l+1}\sum_{m=0}^{l}\frac{\gamma'_{l,k}\gamma_{l,m}}{(2t_>)^k(2t_<)^m}\left[(-1)^m\tilde{S}^{(k+m)}(t_+) + (-1)^{l+1}\tilde{S}^{(k+m)}(t_-)\right] \quad (46)$$

where

$$\gamma'_{l,n} = \gamma_{l,n} + 2(n-1)\gamma_{l,n-1} \quad n=1,2,...,l$$
$$\gamma'_{l,0} = \gamma_{l,0} \qquad \gamma'_{l,l+1} = 2l\gamma_{l,l} \quad (47)$$

With Eqs. (41) through (47) we thus have explicit solutions for all our field components. Since we have assumed $\Omega a \ll c$, the charge distribution in a non-conductor may be considered to be time-independent with its density given by $\delta(r-a)\sigma_l(\theta)$ where

$$\sigma_l(\theta) = \frac{Q}{4\pi a^2}\frac{\Lambda_l(\theta)}{\sin(\theta)} \quad (48)$$

This charge density may be assumed constant throughout any changes in the rotation rate of the sphere. Using the orthogonality of the $\Lambda_l$'s (see Eq. (9)), we can express any angular function $\sigma(\theta)$ as an infinite sum over this set of basis functions, with the solution for **E** and **B** given by the corresponding sum over our solutions as given above. We thus have the complete solution for any axially symmetric angular distribution of charge.

Before continuing, we relate our solutions for the $l = 1$ case to the solutions given by [4] and [5] for this case. In both [4] and [5], the solutions given by these authors for $l = 1$ can be reduced to expressions involving the "moments" $M^{(n)}(t)$ of the time function $\tilde{S}^{(0)}(t)$, defined as

$$M^{(n)}(t) = \int_0^t \tau^n \tilde{S}^{(0)}(\tau)d\tau \quad (49)$$

The results obtained by [4] and [5] for $l = 1$ can be readily expressed in terms of the first three moments $M^{(0)}(t)$, $M^{(1)}(t)$, and $M^{(2)}(t)$. One can show by successive integration by parts that these moments are simply obtainable from our serial integrals $\tilde{S}^{(n)}(t)$. For example, $M^{(1)} = t\tilde{S}^{(1)} - \tilde{S}^{(2)}$. With these relations, the results obtained by [4] and [5] for $l = 1$ can be readily converted to our form in terms of the first three functions $\tilde{S}^{(n)}$, and they demonstrate perfect agreement with our expressions, although the initial appearances are very different.

In addition, with the help of Eqs. (43), (45) and (46) one can show that at the boundary of the spherical shell ($t_< = t_> = t_a$), the expression $[\partial f_l / \partial t_> - \partial f_l / \partial t_<]$ reduces to a single term in our double sums in Eqs.(45) and (46) (the term with $k = m = 0$) and the result satisfies the boundary condition below, as it should. One finds that the $B_\theta$- jump is given by, as we expect,

$$\left[\tilde{B}_{l,\theta}\right]_{t_>=t_a+\varepsilon} - \left[\tilde{B}_{l,\theta}\right]_{t_<=t_a-\varepsilon} = \left[\frac{\partial f_l}{\partial t_<}\right]_{t_<=t_a-\varepsilon} - \left[\frac{\partial f_l}{\partial t_>}\right]_{t_>=t_a+\varepsilon} = \tilde{S}^{(0)}(t) \qquad (50)$$

and this justifies our choice of $C(s)$ given in Eq. (21)

Finally, we summarize the solutions for the situation when the sphere is spinning at *constant* rate, as we shall need them in the discussions that follow. The solutions for this case are well known [9] and we simply quote the results:

$$B_{r,l}(r,\theta) = \frac{l(l+1)}{(2l+1)} P_l(\cos\theta) B_0 \begin{cases} \left(\dfrac{r}{a}\right)^{l-1} & r < a \\ \left(\dfrac{a}{r}\right)^{l+2} & r > a \end{cases} \qquad (51)$$

and

$$B_{\theta,l}(r,\theta) = \frac{\Lambda_l(\cos\theta) B_o}{(2l+1)} \begin{cases} -(l+1)\left(\dfrac{r}{a}\right)^{l-1} & r < a \\ l\left(\dfrac{a}{r}\right)^{l+2} & r > a \end{cases} \qquad (52)$$

For use in our considerations in Section 5 through 7 below, we note that if we use the normalization for the Legendre polynomials,

$$\int_{-1}^{1} P_l^2(x) dx = \frac{2}{2l+1} \qquad (53)$$

and the corresponding normalization for $\Lambda_l$ given in Eq. (10), we can use Eqs. (51) and (52) to compute the total static magnetic energy for the $l^{th}$ multipole as

$$U_l^{mag} = \frac{l(l+1)}{(2l+1)^2} \frac{2\pi a^3 B_o^2}{\mu_o} \tag{54}$$

The ratio of the energy stored inside the sphere to the total energy stored is $(l+1)/(2l+1)$, so that in the limit of large $l$, there are equal amounts of energy stored inside and outside the sphere.

## 5    Numerical examples

We now consider numerical solutions to Eq. (15) for various angular distributions of the charge distribution on the sphere and for various types of time behavior for $\tilde{S}^{(0)}(t)$. We will concentrate on the $l = 1$ case, since that is the most familiar, but will consider other values of $l$ and make general comments about their properties as we proceed. For $l = 1$ the solution to Eq. (7) is $\Lambda_1 = \sin\theta$. The time independent solutions for the electric field are $E_\theta = 0$ and

$$E_r = \frac{Q}{4\pi\varepsilon_o a^2} \begin{cases} \left(\dfrac{a}{r}\right)^2 & \text{for } r \geq a \\ 0 & \text{for } r < a \end{cases} \tag{55}$$

The general expression for $\tilde{E}_{1,\phi}(t_r, t_a, t)$ (see Eq. (41) and Eq.(44)) has four terms, as follows

$$\tilde{E}_{1,\phi}(t_r, t_a, t) = -\frac{t_a}{2t_r} \left[ \begin{array}{l} \left( \left[\tilde{S}^{(0)}(t_+) + \tilde{S}^{(0)}(t_-)\right] - \dfrac{1}{t_<}\left[\tilde{S}^{(1)}(t_+) - \tilde{S}^{(1)}(t_-)\right] \right) \\ + \dfrac{1}{t_>}\left( \left[\tilde{S}^{(1)}(t_+) + \tilde{S}^{(1)}(t_-)\right] - \dfrac{1}{t_<}\left[\tilde{S}^{(2)}(t_+) - \tilde{S}^{(2)}(t_-)\right] \right) \end{array} \right] \tag{56}$$

If we write a similar expression for $\tilde{B}_{1,r}$ [see Eq.(42) and Eq.(30) ), we again have four terms, but now involving $\tilde{S}^{(1)}(t)$ through $\tilde{S}^{(3)}(t)$ instead of $\tilde{S}^{(0)}(t)$ through $\tilde{S}^{(2)}(t)$, as in Eq.(56). The expression for $\tilde{B}_{1,\theta}(t_r, t_a, t)$ (see Eqs.(43), (45) and (46)) has six terms involving $\tilde{S}^{(0)}(t)$ through $\tilde{S}^{(3)}(t)$. In the general $l$ case, $\tilde{E}_{l,\phi}$ is a sum of $(l+1)^2$ terms involving $\tilde{S}^{(0)}(t)$ through $\tilde{S}^{(2l)}(t)$, $\tilde{B}_{1,r}$ has the same number of terms, except they involve $\tilde{S}^{(1)}(t)$ through $\tilde{S}^{(2l+1)}(t)$, and $\tilde{B}_{l,\theta}$ will have $(l+1)(l+2)$ terms involving $\tilde{S}^{(0)}(t)$ through $\tilde{S}^{(2l+1)}(t)$.

To calculate numerical solutions, we must specify $\tilde{S}^{(0)}(t)$. We can specify *any* time dependence that we desire, even a function $\Omega(t) = \Omega_o \tilde{S}^{(0)}(t)$ which is discontinuous and which has discontinuities and/or delta functions in its time derivatives! This is because our general solution in Eq. (30) involves only the rotation rate $\Omega(t)$ and its time *integrals*, and does not depend on the time derivatives of the rotation rate. However, to touch base with the limits with which we are familiar (radiation in the electric dipole approximation), we first choose a time behavior which has continuous derivatives to any order, so that we can check our general solutions with the classic electric dipole formulas for radiation rates in the electric dipole approximation.

We thus initially spin our sphere up smoothly with a characteristic time $T$ using the "smooth" turn-on function

$$\tilde{S}^{(0)}_{smooth}(t) = \frac{1}{2}\left[\frac{2}{\pi}\arctan\left(\frac{t}{T/5}\right) + 1\right] \tag{57}$$

We use this expression to calculate our field components, using Eq.(28) to find $\tilde{S}^{(n)}_{smooth}(t)$ (which can be obtained in analytic form). We first choose a value of $T = 5t_a$. In Figure 1, we plot the radial profiles of $r\tilde{B}_{1,\theta}$ for the dipole ($l = 1$) term and of $r\tilde{B}_{2,\theta}$ for the quadrupole ($l = 2$) term at a time $t = 15t_a$ after a spin-up centered at $t = 0$. We have multiplied the field components by $r$ in this plot to bring out clearly the radiation term behavior. These radiation terms decrease as inverse $r$, where as all other terms decrease faster than $1/r$. The vertical scale in Figure 1 is arbitrary, but linear.

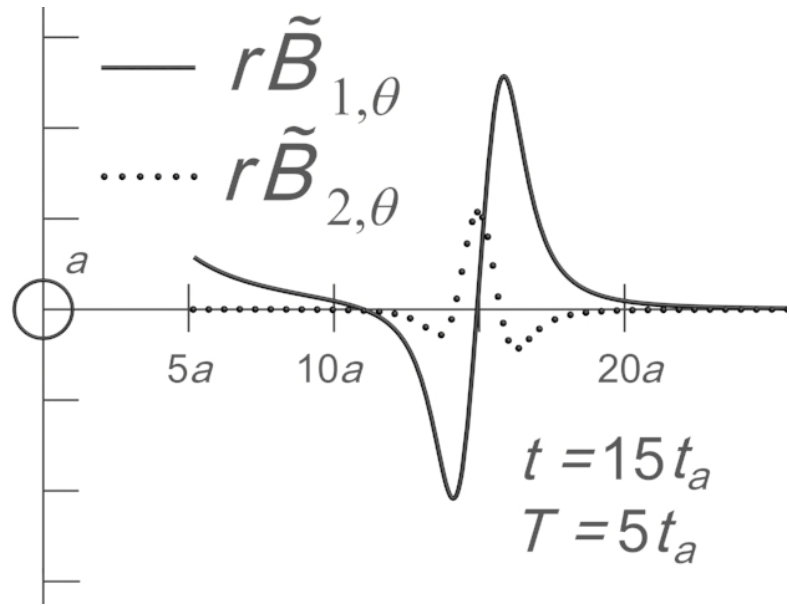

**Figure 1:** The dipole $l = 1$ and quadrupole $l = 2$ field components $r\tilde{B}_\theta$ at a time $t = 15t_a$ after a "smooth" spin-up centered at $t = 0$ over a time $T$, with $T = 5t_a$.

The electric dipole approximation is $t_a/T = a/cT \ll 1$, and we have a value of $t_a/T = 1/5$ in Figure 1. We see all of the qualitative characteristics we expect to see in the electric dipole approximation. The magnetic dipole term $\tilde{B}_{1,\theta}$ shows a time behavior in its radiation fields which is proportional to the second time derivative of $\Omega(t)$, and the magnetic quadrupole term shows a behavior in its radiation fields which is proportional to the third time derivative of $\Omega(t)$ (cf. Eq. (62) of Section 6 below). The amplitude of the quadrupole radiation field is significantly smaller than that of the dipole radiation field, as expected (in the electric dipole approximation it should be down by a factor of $t_a/T \ll 1$).

In contrast, we now consider situations in which the electric dipole approximation is clearly invalid. In Figure 2 we plot the radial profiles of $\tilde{B}_{1,\theta}$ and $\tilde{B}_{2,\theta}$ for a much faster spin-up time, $T = t_a/10$, at a time $t = 3t_a$. This short time scale grossly violates the electric dipole approximation ($t_a/T = 10$, hardly small compared to 1). In Figure 2 we no longer multiply the $\tilde{B}_\theta$ components by $r$ to bring out the radiation term behavior, since the radiation terms are so much larger with this shorter spin-up time. The vertical scale in Figures 2 is now absolute (but dimensionless), so that we can directly compare our numerical results with our theoretical expectations for the field amplitudes.

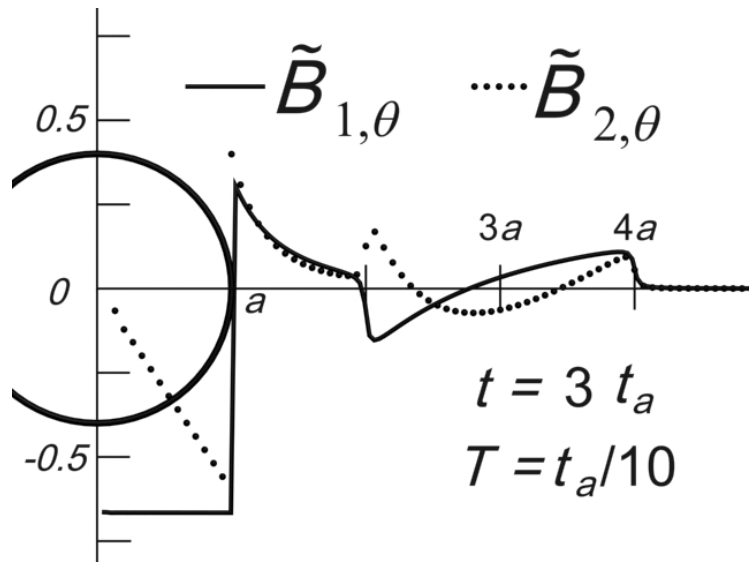

**Figure 2:** The dipole $l = 1$ and quadrupole $l = 2$ $\tilde{B}_\theta$ field components at a time $t = 3t_a$ after a "smooth" spin-up centered at $t = 0$ over a time $T$, with $T = t_a/10$.

We see in Figure 2 many of the characteristics we would expect for this short spin-up time, and many that we would perhaps not expect. The radiation fields for $t > 2t_a$, as in this figure, are entirely outside the sphere and confined within a radial extent of $2a$ (more closely $2a + cT$, but in the limit of $T \ll t_a$ this is approximately $2a$). For $t > 2t_a$, the dipole radiation fields ($l = 1$) have one zero in this $2a$ radial extent, the quadrupole radiation fields ($l = 2$) have two zeroes, and in general the *l*-th multipole radiation term has *l* zeroes in this interval. The amplitude of the quadrupole radiation field is very similar to that of the dipole radiation field.

To give a feel for what the zeroes in the radiation fields in Figure 2 mean in terms of the overall topology of the field, we show in Figure 3 a line integral convolution representation of the $l = 2$ magnetic field at the same time as the radial profile for $\tilde{B}_{2,\theta}$ given in Figure 2. We have color-coded this figure so that the red zones correspond to fields generated when the sphere is in the process of spinning up, the darker blue zones are the radiation fields generated while the sphere is at constant speed, and the lighter blue zones are the static fields. The two zeroes in the $l = 2$ radiation profile in the interval $2a < r < 4a$ imply that there are two "lobes" in the radiation field structure in radius, just as there are two lobes in polar angle for $l = 2$. For the *l*-th multipole case there will be *l* lobes in both radius and polar angle. Examples of this behavior can be found at ://web.mit.edu/viz/spin.

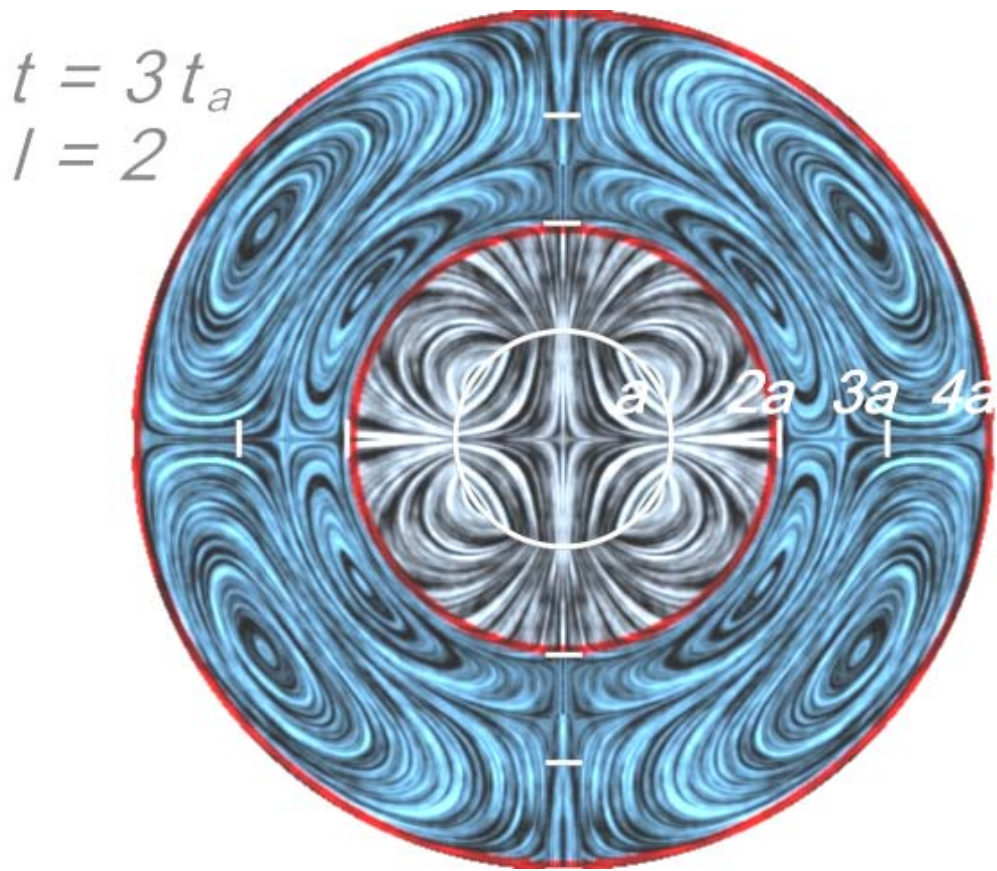

**Figure 3:** A line integral convolution representation of the $l = 2$ magnetic field configuration at a time $t = 3t_a$, after a rapid spin-up at $t = 0$, corresponding to the radial profile for $\tilde{B}_{2,\theta}$ in Figure 2.

Now let us look at much earlier times, just after we have spun-up the sphere very rapidly. In Figure 4, we again have the very short spin-up time $T = t_a/10$ of Figure 2 and 3, but we are now looking at a much earlier time, $t = 0.4 t_a$, a time short compared to the $2t_a$ speed of light transit time across the sphere. Now both $\tilde{B}_{1,\theta}$ and $\tilde{B}_{2,\theta}$ have very similar profiles, and they are both what we would expect for the fields of an infinite plane accelerated abruptly, as given in Eq. (1), as one would expect.

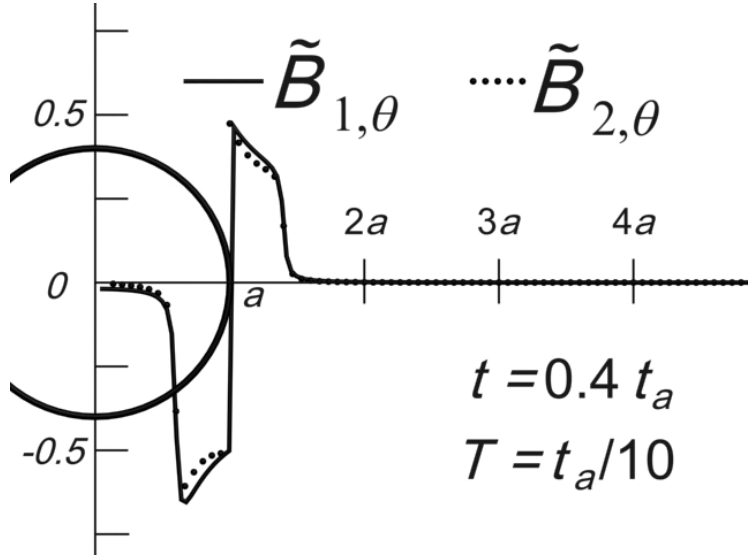

**Figure 4:** The dipole $l = 1$ and quadrupole $l = 2$ $\tilde{B}_\theta$ field components at a time $t = 0.3 t_a$ after a "smooth" spin-up centered at $t = 0$ over a time $T$, with $T = t_a / 25$.

Finally, we can numerically compute the amount of energy radiated away in our spin-up process. Using Eqs.(41) and (43), we have the following general expression for the radial component of the Poynting flux **S** for $t_r \geq t_a$

$$S_{l,r} = -\frac{E_{l,\phi} B_{l,\theta}}{\mu_o} = -\frac{cB_0^2}{\mu_o}\Lambda_l^2(\theta)\left(\frac{t_a}{t_r}\right)^2 \frac{\partial f_l(t_>,t_<,t)}{\partial t} \frac{\partial f_l(t_>,t_<,t)}{\partial t_>} \tag{58}$$

If we are interested in the total energy radiated, we need only keep the inverse radius squared terms in Eq.(58), and integrate over time and surface area of a sphere at infinity. For example, for the $l = 1$ case, using the "smooth" spin-up behavior given in Eq. (57), we can numerically compute an expression for the total energy radiated during the spin-up for any value of $t_a / T$. This energy, normalized to $U_1^{mag}$ (the energy stored in the $l = 1$ static magnetic field (see Eq.(54)), is plotted as a function of the ratio $t_a / T$ in Figure 5, and labeled "Smooth". For comparison, we also plot in Figure 5 the straight line labeled "Dipole". This is the normalized radiated energy calculated from the usual expression for the magnetic dipole radiation rate in the electric dipole approximation, $\mu_o \ddot{m}^2 / 6\pi$ [10]. At low values of the ratio $t_a/T$, our "smooth" solution for the energy radiated in the $l = 1$ mode varies as $(t_a/T)^3$, and is essentially identical with the classical result for magnetic dipole radiation. However, as the ratio $t_a/T$ becomes comparable to and much greater than unity, our numerical result for the energy radiated in the $l = 1$ case in the "smooth" case approaches the constant value $U_1^{mag}$, whereas the "Dipole" approximation formula increases without limit. That is, the energy radiated for our general "smooth" solution in this limit is equal to the energy stored in the static field after spin-up for the $l = 1$ case.

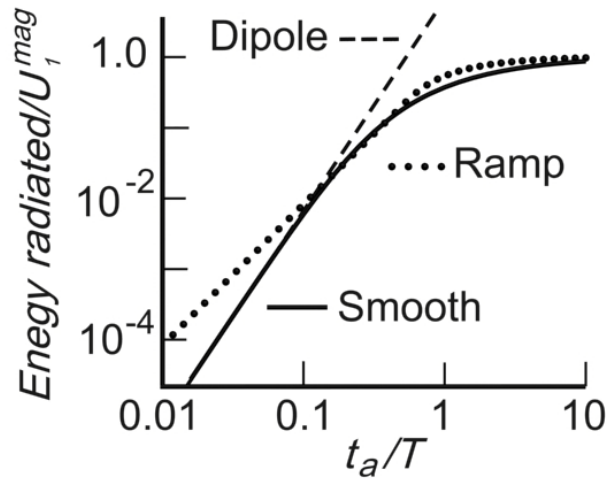

**Figure 5:** The dipole $l = 1$ normalized energy radiated as a function of $t_a/T$ for the smooth spin-up with characteristic time $T$, and also for a ramp spin-up over time $T$. The curve labeled *Dipole* is the electric dipole approximation for the normalized radiated energy for the smooth spin-up time function.

The remarkable thing about our general solutions is that we are free to consider any less well-behaved function for the spin time dependence, since our general solutions make no assumption about the "niceness" of the spin-up function. For example, we show in Figure 5 the total radiated energy versus $t_a/T$ for a "ramp" spin-up, that is a function $\tilde{S}_{ramp}^{(0)}(t)$ which is zero for $t < 0$ and 1 for $t > T$, and increases linearly from 0 to 1 in the interval $0 \leq t \leq T$. This function is, of course, not expandable in a Taylor series, and the classic expression for the radiated power in the electric dipole limit, proportional to the square of the second time derivative of the magnetic dipole moment, cannot be evaluated. However there is no problem in evaluating our expressions in Eq.(58) using $\tilde{S}_{ramp}^{(0)}(t)$ to compute the fields, and we show that numerical calculation for the ramp spin-up in Figure 5. The behavior at low values of $t_a/T$ is now proportional to $(t_a/T)^2$, and this is understandable analytically (with some effort). The behavior at high values of $t_a/T$ is the same as for the smooth function, approaching the same constant, as we expect, since in this limit both $\tilde{S}^{(0)}_{ramp}(t)$ and $\tilde{S}^{(0)}_{smooth}(t)$ approach a unit step function at $t = 0$.

## 6    Recovering the electric dipole approximation solutions

We now return to our analytic efforts, and show explicitly that we recover the electric dipole approximation solutions for the radiation fields from our general form, as we would expect given the numerical solutions we have shown above in Figures 1 and 5. Recall that electric dipole approximation implies that the radiating sources are confined to such a small region that $t_a = a/c$ is negligible compared to the spin-up time $T$. To recover the classic electric dipole radiation terms, we make the assumption (unnecessary in our general solution) that $\Omega(t)$ is analytical, *i.e.,* that it and all of its derivatives are well behaved, so that it can be expanded in a Taylor series as follows:

$$g(t-t_> \pm t_<) = \sum_{n=0}^{\infty} \frac{(\pm t_<)^n}{n!} \frac{d^n g(t-t_>)}{dt^n} \tag{59}$$

If we examine our solutions for $f_l$ in Eq.(30), assume that $t_r > t_a$, and keep only the $k = 0$ term (which will give the radiation fields), and expand each $\tilde{S}^{(m)}(t-t_r \pm t_a)$ using Eq.(59), we can show that $f_l^{rad}$, the radiation part of $f_l$, is given by

$$f_l^{rad} = \frac{1}{2} \sum_{n=0}^{\infty} \frac{t_a^n}{n!} \left\{ \sum_{m=0}^{l} \frac{\gamma_{l,m}}{(-2t_a)^m} \left[ 1 + (-1)^{l+1+n-m} \right] \tilde{S}^{(m-n+1)}(t-t_r) \right\} \tag{60}$$

where $\tilde{S}^{(-n)}(t) = d^n \tilde{S}^{(0)}(t)/dt^n$. Since the behavior of $f_l^{rad}$ near the origin of the sphere is characterized by the behavior of the modified spherical Riccati-Bessel functions (as contrasted with modified spherical Hankel functions), we expect $f_l^{rad}$ to go to zero as $r^{l+1}$. And this is the behavior we find, in that the first non-zero term in Eq. (60) does not occur until the index $j$ in $\tilde{S}^{(j)}$ is equal to $-l$. That is, the first non-zero term in Eq.(60) is proportional to the *l-th* derivative of $\tilde{S}^{(0)}$, and therefore the radiation **E** and **B** fields are proportional to the *(l+1)-th* time derivative of $\tilde{S}^{(0)}(t)$. Thus, for $l = 1$ the radiation fields are proportional to the second time derivative, and for $l = 2$ the third time derivative, and so on, just as we find in the usual electric dipole expansion. If we look at the first non-zero term in Eq.(60), for which $m - n + 1 = -l$, or $n = l + m + 1$, it is

$$f_l^{rad} = t_a^{l+1} \tilde{S}^{(-l)}(t-t_r) \sum_{m=0}^{l} \left( -\frac{1}{2} \right)^m \frac{\gamma_{l,m}}{(l+m+1)!} \tag{61}$$

This expression can be written in closed form by using the binomial theorem and the beta function (see Appendix)

$$f_l^{rad} = \frac{l!}{2(2l+1)!}(2t_a)^{l+1}\tilde{S}^{(-l)}(t-t_r) \equiv \frac{t_a^{l+1}}{1 \cdot 3 \cdot 5 \cdots (2l+1)}\tilde{S}^{(-l)}(t-t_r) \tag{62}$$

We can derive the amplitude of the radiation fields for the *l-th* multi-pole in the electric dipole approximation using Eq. (62). These expressions agree with the classic expressions that we find in the literature for $l = 1$ and $l = 2$. We expect them to agree for all orders of $l$.

## 7      The "instantaneous" radiation fields and the energy radiated

Given the results in Figure 5 for the total energy radiated as a function of $(t_a/T)$ for $l = 1$, we might suspect that for any $l$ the energy radiated away for instantaneous turn on is equal to the energy stored in the long term magnetostatic field for that $l$. Because of the orthogonality of the various angular functions for different $l$, if this is true for each $l$, it is also true for a linear sum of $l$'s. Since we can expand any distribution of charge using the complete set of functions given by the $\Lambda_l$'s, this would mean that the abrupt spin-up of any axially symmetric distribution of charge on the sphere will radiate away exactly the same amount of energy as stored in the resultant magnetostatic field after spin-up.

This is indeed the case, as we now argue. Let us consider the "instantaneous" turn-on solutions. That is, we evaluate the electric and magnetic fields radiation fields when $\Omega(t)$ goes from zero to a constant value in zero time. In this case

$$\tilde{S}^{(n)}(t) = \frac{t^n h(t)}{n!} \tag{63}$$

where $h(t)$ is the unit step function

$$h(t) = \begin{cases} 1 & \text{for } t \geq 0 \\ 0 & \text{otherwise} \end{cases} \tag{64}$$

If we assume $t_r > t_a$ and confine our attention to the asymptotic radiation fields only, ($r \to \infty$), we may use the approximation $\partial f_l / \partial t_> = \partial f_l / \partial t_r \cong -\partial f_l / \partial t$ and retain only the zero-order terms in the $(1/t_r)$-polynomial in Eq.(58). This implies that we need to evaluate only one function to get the Poynting flux at infinity, $\partial f_l / \partial t = -t_r \tilde{E}_\phi / t_a$. After some effort, we find

$$\frac{\partial f_l^{rad}}{\partial t} = \frac{1}{2_a} \sum_{m=0}^{l} (2t_a)^{-m} \gamma_{l,m} \left[ (-1)^m \tilde{S}^{(m)}(t_+) + (-1)^{l+1} \tilde{S}^{(m)}(t_-) \right]$$

$$= \frac{1}{2_a} \sum_{m=0}^{l} \frac{\gamma_{l,m}}{m!} \left[ \left( -\frac{t_+}{2t_a} \right)^m h(t_+) + (-1)^{l+1} \left( \frac{t_-}{2t_a} \right)^m h(t_-) \right] \qquad (65)$$

where, according to Eq.(31),

$$t_\pm = t - t_r \pm t_a \qquad (66)$$

Some thought shows that the expression in Eq.(65) is zero for $t_+ \leq 0$ and for $t_+ > 2t_a$, which implies that we have simply

$$\frac{\partial f_l^{rad}}{\partial t} = -\frac{\partial f_l^{rad}}{\partial t_r} = \begin{cases} \frac{1}{2} \sum_{m=0}^{l} \frac{\gamma_{l,m}}{m!} \left( -\frac{t_+}{2t_a} \right)^m & \text{if } 0 < t_+ \leq 2t_a \\ 0 & \text{otherwise} \end{cases} \qquad (67)$$

Note that, according to Eq.(67), the radiation fields will vary as a polynomial in $(t - t_r + t_a)$ up to the $l^{th}$ power in the interval $0 \leq t - t_r + t_a < 2t_a$. There will be $l$ zeroes of this polynomial within this interval, each zero adding an additional "lobe" in radius (see the two lobes in Figure 3 for $l = 2$, for example).

If we insert the expression for the instantaneous spin-up fields in Eq. (67) into the general expression for the Poynting flux in Eq. (58), we find that $S_{l,r}$ is zero except in the interval $0 < t_+ < 2t_a$ and in that interval it is given by

$$S_{l,r}(r,\theta) = \frac{cB_0^2}{\mu_o} \Lambda_l^2(\theta) \left( \frac{t_a}{t_r} \right)^2 \left\{ \frac{1}{2} \sum_{m=0}^{l} \frac{\gamma_{l,m}}{m!} \left( \frac{-t_+}{2t_a} \right)^m \right\}^2 \qquad (68)$$

Using this expression, we can compute the total energy radiated into all solid angles. Integrating Eq. (68) over the area of a sphere at infinity and over time in the interval $0 < t_+ < 2t_a$ gives for the total energy radiated in the $l^{th}$ multipole

$$U_l^{rad} = \frac{l(l+1)}{2l+1} \frac{2\pi a^3 B_o^2}{\mu_o} \int_0^1 \left\{ \frac{1}{2} \sum_{m=0}^{l} \frac{\gamma_{l,m}}{m!} (-\eta)^m \right\}^2 d\eta \qquad (69)$$

But we can show that (see the Appendix)

$$\int_0^1 \left\{ \sum_{m=0}^{l} \frac{\gamma_{l,m}}{m!} (-\eta)^m \right\}^2 d\eta = \frac{1}{(2l+1)} \qquad (70)$$

Comparing Eq.s (69) and (70) with the expression for $U_l^{mag}$ in Eq. (54), we see that the radiated energy is equal to the energy stored in the abrupt spin-up case.

This fact is also consistent with an evaluation of the work done in spinning up the sphere instantaneously. We calculate this work by evaluating the integral of $-\mathbf{E} \cdot \mathbf{J}$. We only need the electric field at the surface of the sphere as a function of time, since it is only there that the current is non-zero. After some effort one finds that the work $U_l^{work}$ done in spinning up the sphere instantaneously for the $l^{th}$ multipole is given by

$$\frac{U_l^{work}}{2U_l^{mag}} = (2l+1) \sum_{k=0}^{l} \sum_{m=0}^{l} \frac{(-1)^m \gamma_{lk} \gamma_{lm}}{(k+m+1)!} \qquad (71)$$

The double sum in Eq. (71) is unity (see Appendix). Thus the work done in spinning up the sphere instantaneously is just twice the energy stored, or the sum of the energy stored plus the energy radiated, as it must be.

## 8   Magnetic field lines

Our geometry permits an easy integration of the ordinary differential equation describing the pattern of the force lines of the magnetic field at a given instant in the meridian plane of the spherical shell. The integration is elementary and, in principle, can be carried out analytically for any $l$. We present here three explicit formulas for a field line at a fixed time $t$. The differential equation for a field line according to Eq.'s (42), (43), (44), (45), and (46), is

$$\frac{rd\theta}{dr} == \frac{B_\theta(r,\theta,t)}{B_r(r,\theta,t)} = -\frac{\Lambda_l(\theta) \, t_r}{l(l+1) P_l(\cos\theta) f_l(t_r,t)} \frac{\partial f_l(t_r,t)}{\partial t_r} \qquad (72)$$

If we change the polar angle variable $\theta$ to $x = \cos\theta$, and use Eq. (8), we can write Eq. (72) as

$$\frac{\partial f_l(t_r,t)}{\partial t_r} \frac{dt_r}{f_l(t_r,t)} - \frac{l(l+1) P_l(x) dx}{(1-x^2)(dP_l(x)/dx)} = 0 \qquad (73)$$

Integrating we get

$$\ln[f_l(r,t)] - \Phi(x) = const \qquad (74)$$

where we put for short

$$\Phi(x) = \int \Phi'_l(x)dx; \quad \Phi'_l(x) = \frac{l(l+1)P_l(x)}{(1-x^2)P'_l(x)} \tag{75}$$

We have, in succession,

$$\Phi'_1 = \frac{2x}{1-x^2}; \quad \Phi'_2 = \frac{3x^2-1}{(1-x^2)x}; \quad \Phi'_3 = \frac{4x(5x^2-3)}{(1-x^2)(5x^2-1)} \tag{76}$$

The integrations over $x$ are elementary. One finds for the first three $l$'s

$$\begin{aligned}\Phi_1 &= -\ln(1-x^2)\\ \Phi_2 &= -2\ln\left[x(1-x^2)\right]\\ \Phi_3 &= -\ln\left[(1-x^2)(x^2-\tfrac{1}{5})\right]\end{aligned} \tag{77}$$

which, in terms of $\theta$, yields for a field line at a fixed time $t$:

$$\begin{aligned} f_1(r,t)\sin^2\theta &= f_1(r_o,t)\sin^2\theta_o\\ f_2(r,t)\cos^2\theta\sin^4\theta &= f_2(r_o,t)\cos^2\theta_o\sin^4\theta_o\\ f_3(r,t)\sin^2\theta\left(\cos^2\theta-\tfrac{1}{5}\right) &= f_3(r_o,t)\sin^2\theta_o\left(\cos^2\theta_o-\tfrac{1}{5}\right)\end{aligned} \tag{78}$$

The field point $(r_o,\theta_o)$ is a *reference* point through which the field line passes at $t$.

## 9    Acceleration without radiation

We finally consider a very different time behavior for $\tilde{S}^{(0)}(t)$ that offers insight into the "acceleration without radiation" considerations in [2]. Let $\tilde{S}^{(0)}(t)$ be zero for $t < 0$, and for $t > 0$ suppose $\tilde{S}^{(0)}(t)$ oscillates sinusoidally at frequency $\omega$. In Figure 6, we show the radial profile of $B_{1,\theta}$ and $E_{1,\phi}$ in the equatorial plane for the $l = 1$ mode. These radial profiles are shown at a time $3.0\ t_a$ after the sphere begins to oscillate. We have taken the oscillation frequency $\omega$ to be such that $\dfrac{\omega a}{c}$ is the first zero of $j_1(x)$. If we look our solutions in Eq. (20) for a purely harmonic time series, we see that the field outside $r = a$ is zero for a pure oscillation at this frequency. For the semi-harmonic time behavior we show in Figure 6, we see this purely harmonic behavior establish itself after a time $2t_a$ from the beginning of the oscillation, as we might expect, accompanied by a burst of radiation in the interval from 0 to $2t_a$. This sequence shows how the "acceleration without radiation" situation considered in [2] establishes itself in the time domain. We conjecture that the energy radiated away in establishing this situation is a constant fraction of the energy stored in the standing waves within the sphere for $t > 2t_a$, but we have not shown this.

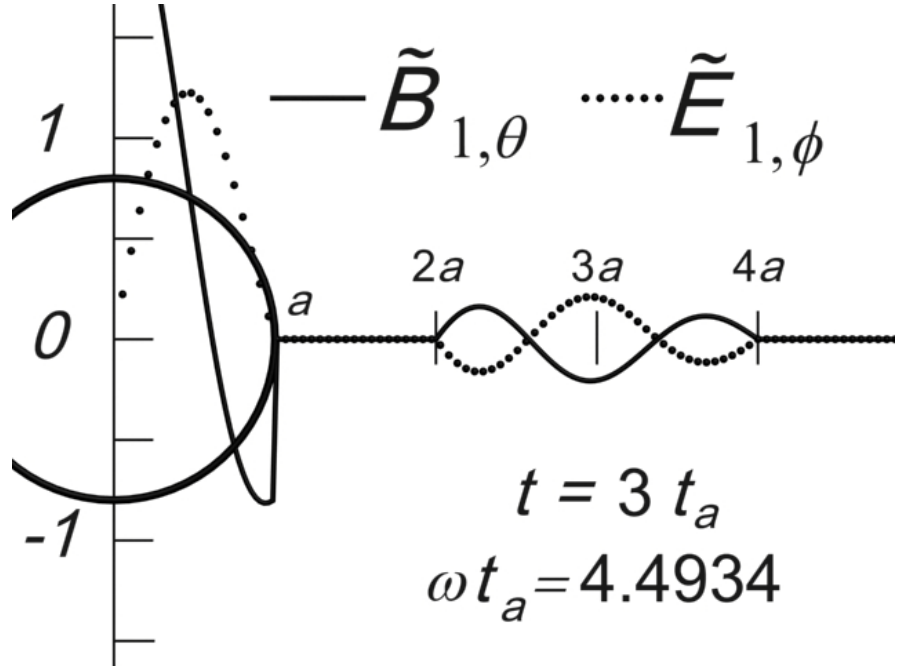

**Figure 6:** The dipole $l = 1$ fields in the equatorial plane at a time $t = 3t_a$ for a semi-sinusoidal oscillation starting at $t = 0$, where the oscillation frequency $\omega$ is such that $\omega a / c$ it is the first zero of $j_1(x)$.

## 10  Discussion and Conclusions

    We have considered the problem of spinning up a charged spherical shell of radius $a$ carrying an arbitrary axially symmetric distribution of charge. We have found that the general solution to this problem can be written in terms of a sum over multipole solutions. These multipole solutions involve the rotation rate and its time integrals, with the solution for the *l-th* multipole involving integrals of the rotation rate up to its *(2l+1)-th* time integral, as shown in Eq. (46). Surprisingly, our solutions do *not* depend on the time derivatives of the rotation rate. This is in contrast to the classic solutions in the electric dipole approximation, where the *l-th* mutipole radiation fields depend on the *(l+1)-th* time derivative of the rotation rate (see Eq (62)). We have shown that our solutions in the general case reduce to those of the electric dipole approximation when the spin-up time period is large compared to $a/c$, as they must.

    We have also presented numerical solutions for various distributions of surface charge. We have paid particular attention to the case where the sphere instantaneously goes from rest to rotating at a constant rate. For a time $2a/c$ after this instantaneously spin-up, the sphere radiates, even though it is rotating at a constant rate. For such abrupt spin-ups, we have shown that the sphere radiates away an amount of energy exactly equal

to the magnetic energy stored in the magnetic field a long time after the sphere has been spun up.

Our general solutions give some insight into the "radiation without acceleration" considerations of [2], in that we can see how those non-radiating solutions develop in the time domain. Using our general formalism, we can see how solutions which initially radiate evolve to solutions which do not radiate because of destructive interference from different parts of the finite charge distribution. Animations of such solutions, as well as of other solutions, are available at ://web.mit.edu/viz/spin .

In conclusion, we emphasize three points. The first is that the classic electric dipole radiation approximation for the spinning sphere takes the sphere to be "point-like". In our general solutions to this problem, we have relaxed that assumption, allowing the sphere to have any finite radius $a$ compared to $cT$. We have also looked in detail at the properties of our general solutions in essentially the opposite extreme of the electric dipole approximation. That is, we have in part focused on the situation where the radius of the sphere goes to infinity for a given time scale $T$, that is, on the case that $a >> cT$. We find a number of novel properties of this "instantaneous spin-up", most notably a characteristic signature in the profile of the radiation fields given by Eq. (67), and the equality of energy radiated and energy stored.

The second point we emphasize is the speed of calculation. In our solutions we do not have to Fourier analyze in the time domain, go to the frequency domain, and then return to the time domain, a problematic procedure in times of computing time. We are able to write down solutions in the time domain without ever analyzing our source function in the frequency domain. Thus although in principle our solutions are contained in the procedure outlined in [3], in practice it would be prohibitively difficult to use that procedure to find time domain solutions. In contrast, our solutions, at least for lower order multipoles, can be computed rapidly, even in the limit of extreme time behaviors (e.g. step functions in time).

The final point we emphasize is the novelty of our solutions when compared to the classic electric dipole ones, even in the limit that $a << cT$. In many aspects classical electromagnetism deals with the "physics of singularities", and it is striking how different our solutions are when we are able to move away from singular structures. Most notably, our solutions for non-singular spatial structures involve the serial integrals of a function of time, not serial differentiations of that time function. Although we recover the differential behavior from our solutions when we go to the singular limit, there is no fundamental requirement that we have a time function that is continuous and differentiable. Indeed we are able to address problems even in the limit that $a << cT$ which we could never address in the electric dipole approximation, e.g. dipole moments which are discontinuous in time or have discontinuities or delta functions in their time derivatives.

It is this last aspect of our solutions that we find to be the most surprising.

## 11 Appendix

First we outline the proof of Eq. (70). We recall that the beta function is defined as [8]

$$B(m+1, l+1) = \int_0^1 \eta^m (1-\eta)^l \, d\eta = \frac{m! \, l!}{(m+l+1)!} \tag{79}$$

Let $\psi_l^{(0)}(\eta)$ be defined by (cf. Eq. (24))

$$\psi_l^{(0)}(\eta) = \sum_{m=0}^{l} \frac{\gamma_{l,m}}{m!}(-\eta)^m = \sum_{m=0}^{l} \frac{(l+m)!}{(m!)^2 (l-m)!}(-\eta)^m \tag{80}$$

and define the successive integrals of $\psi_l^{(0)}(\eta)$ by

$$\psi_l^{(n)}(\eta) = \int_0^\eta \psi_l^{(n-1)}(\eta') \, d\eta' \qquad n = 1, 2, 3 \ldots, l \tag{81}$$

and, in a reverse procedure, its $(2l+1)$ successive derivatives by

$$\psi_l^{(l-n)}(\eta) = \frac{d^n}{d\eta^n} \psi_l^{(l)} \qquad n = 0, 1, 2, 3 \ldots, l \tag{82}$$

and

$$\psi_l^{(-n)}(\eta) = \frac{d^n}{d\eta^n} \psi_l^{(0)} \qquad n = 1, 2, 3 \ldots, l \tag{83}$$

With these definitions, Eq. (70) becomes

$$\int_0^1 \{\psi_l^{(0)}(\eta)\}^2 \, d\eta = \frac{1}{(2l+1)} \tag{84}$$

To prove that Eq. (84) is true, we first easily show by integrating Eq. (80) $l$ times that

$$\psi_l^{(l)}(\eta) = \frac{\eta^l}{l!} \sum_{m=0}^{l} \frac{l!(-\eta)^m}{m!(l-m)!} = \frac{\eta^l}{l!}(1-\eta)^l \tag{85}$$

where we have used the binomial theorem to arrive at the final expression in Eq. (85). By taking successive derivatives of the final expression in Eq. (85), we recover an alternate expansion for $\psi_l^{(0)}(\eta)$ as compared to that in Eq. (80), namely

$$\psi_l^{(0)}(\eta) = \sum_{m=0}^{l} \left( \frac{l!}{m!(l-m)!} \right)^2 (-\eta)^{l-m} (1-\eta)^m \qquad (86)$$

We thus have

$$\psi_l^{(n)}(0) = \psi_l^{(n)}(1) = 0 \qquad n = 1, 2, 3..., l \qquad (87)$$

and

$$\psi_l^{(0)}(0) = 1; \quad \psi_l^{(0)}(1) = (-1)^l \qquad (88)$$

We also easily have that $\psi_l^{(-l)}(\eta)$ is independent of $\eta$ and given by

$$\psi_l^{(-l)}(\eta) = (-1)^l \frac{(2l)!}{l!} \qquad (89)$$

If we then integrate Eq. (84) $l$ times by parts we have

$$\int_0^1 \{\psi_l^{(0)}(\eta)\}^2 d\eta = (-1)^l \int_0^1 \psi_l^{(-l)} \psi_l^{(l)} d\eta = \frac{(2l)!}{(l!)^2} \int_0^1 \eta^l (1-\eta)^l d\eta \qquad (90)$$

where we have used Eq. (85) and (89) to arrive at the final expression in Eq. (90). The integral in the last term in Eq. (90) is $B(l+1, l+1) = (l!)^2 / (2l+1)!$. Using this in Eq. (90) yields Eq. (84), as desired.

We use a similar procedure to prove that the double sum in Eq. (71) is unity. To do this, we use Eq. (79) to write

$$\frac{1}{(k+m+1)!} = \frac{1}{k!m!} \int_0^1 \eta^m (1-\eta)^k d\eta \qquad (91)$$

Inserting Eq. (91) into Eq. (71) then gives the following expression for the double sum in Eq. (71).

$$(2l+1) \sum_{k=0}^{l} \sum_{m=0}^{l} \frac{(-1)^m \gamma_{lk} \gamma_{lm}}{k!m!} \int_0^1 \eta^m (1-\eta)^k d\eta = (2l+1) \int_0^1 \psi_m^{(0)}(\eta) \psi_k^{(0)}(\eta-1) d\eta \qquad (92)$$

The integral in Eq. (92) can be evaluated in a manner analogous to the evaluation of Eq. (84), with the same result.